\documentclass[useAMS,usenatbib]{mn2e}
\usepackage{graphicx}
\usepackage{times}

\newcommand{\MSUN}{\mbox{$\mathrm{M_{\odot}}$}}
\newcommand{\MJUP}{\mbox{$\mathrm{M_J}$}}

\title[The secondary eclipse of NN Ser]{Timing variations in the secondary eclipse of NN Ser}

\author[S. G. Parsons et al.]{S.\,G.\,Parsons$^{1}$\thanks{steven.parsons@uv.cl},
T.\,R.\,Marsh$^{2}$,
M.\,C.\,P.\,Bours$^{2}$,
S.\,P.\,Littlefair$^{3}$,
C.\,M.\,Copperwheat$^{4}$,
\newauthor
V.\,S.\,Dhillon$^{3}$,
E.\,Breedt$^{2}$,
C.\,Caceres$^{1,5}$
and M.\,R.\,Schreiber$^{1,5}$
\\
$^{1}$ Departmento de F{\'i}sica y Astronom{\'i}a, Universidad de
Valpara{\'i}so, Avenida Gran Bretana 1111, Valpara{\'i}so 2360102, Chile\\
$^{2}$ Department of Physics, University of Warwick, Coventry CV4 7AL, UK\\
$^{3}$ Department of Physics and Astronomy, University of Sheffield, Sheffield S3 7RH, UK\\
$^{4}$ Astrophysics Research Institute, Liverpool John Moores University, Twelve Quays House, Birkenhead CH41 1LD, UK\\
$^{5}$ Millenium Nucleus "Protoplanetary Disks in ALMA Early Science", Universidad de Valparaiso, Valparaiso 2360102, Chile}
\begin{document}
\input{references.cls}
\date{Accepted 2013 November 19.  Received 2013 November 14; in original form 2013 October 22}

\pagerange{\pageref{firstpage}--\pageref{lastpage}} \pubyear{2013}

\maketitle

\label{firstpage}

\begin{abstract}

The eclipsing white dwarf plus main-sequence binary NN\,Serpentis provides one of the most convincing cases for the existence of circumbinary planets around evolved binaries. The exquisite timing precision provided by the deep eclipse of the white dwarf has revealed complex variations in the eclipse arrival times over the last few decades. These variations have been interpreted as the influence of two planets in orbit around the binary. Recent studies have proved that such a system is dynamically stable over the current lifetime of the binary. However, the existence of such planets is by no means proven and several alternative mechanisms have been proposed that could drive similar variations. One of these is apsidal precession, which causes the eclipse times of eccentric binaries to vary sinusoidally on many year timescales. In this paper we present timing data for the secondary eclipse of NN\,Ser and show that they follow the same trend seen in the primary eclipse times, ruling out apsidal precession as a possible cause for the variations. This result leaves no alternatives to the planetary interpretation for the observed period variations, although we still do not consider their existence as proven. Our data limits the eccentricity of NN\,Ser to $e<10^{-3}$. We also detect a $3.3\pm1.0$ second delay in the arrival times of the secondary eclipses relative to the best planetary model. This delay is consistent with the expected $2.84\pm0.04$ second R{\o}mer delay of the binary, and is the first time this effect has been detected in a white dwarf plus M dwarf system.

\end{abstract}


\section{Introduction}

Recent years have seen the discovery of planets in a number of unusual systems. From the initial discovery of the first exoplanet around the pulsar PSR\,1257+12 \citep{wolszcan92}, planets have also been detected around giant stars \citep{dollinger11}, pulsating stars \citep{collier10,herrero11}, brown dwarfs \citep{chauvin05,han13} and in orbits around binary stars \citep{doyle11,welsh12}. There is now growing evidence for the existence of planets around white dwarfs, both from observations of rocky planetary material being accreted by white dwarfs (see for example \citealt{gansicke12}) and from timing studies \citep{lee09,beuermann10,potter11,beuermann12,marsh13}. These timing studies make use of the small size of the white dwarf which, when in an eclipsing binary system, leads to sharp eclipse features and hence precise times.

The potential existence of planets around these close binaries, which have undergone a common envelope phase of evolution, raises interesting questions about their formation and evolution. The fact that many of the proposed planetary systems would have been unstable prior to the common envelope stage (see for example \citealt{mustill13}) has led to the idea that these planets could instead have formed from the common envelope material itself \citep{beuermann11,veras12}. This idea is also supported by the fact that fewer than 10\% of main-sequence binaries host circumbinary planets, whilst timing data has revealed period variations in virtually every system with long enough coverage \citep{zorotovic13}, implying that the common envelope environment could be conducive to planet formation.

Despite the large number of evolved eclipsing binaries showing variations in the arrival times of their eclipses, none of the circumbinary planets proposed to explain these variations have been confirmed. Indeed, many proposed planetary orbits are ruled out as soon as new data are obtained \citep{parsons10}, whilst many others are dynamically unstable \citep{potter11,hinse12,horner13}. The lack of independent evidence for the existence of these planets call in to question the planetary explanation for the timing variations. Indeed several other mechanisms are able to drive similar variations in the eclipse arrival times such as apsidal motion and fluctuations in the gravitational quadrapole moment of the main-sequence star, known as Applegate's mechanism \citep{applegate92}. However, in some cases the main-sequence stars are unable to provide the required energy to drive the observed period variations via Applegate's mechanism \citep{brinkworth06}.

Apsidal motion is more difficult to rule out since large variations in eclipse arrival times are possible even with very small eccentricities. This is because the amplitude of the timing shifts via apsidal motion is proportional to the orbital period and the eccentricity. Hence an eccentricity of as low as $10^{-3}$ can still drive a deviation in the eclipse times of as large as 10 seconds in a binary with a period of 3 hours. The timescale for this variation can be anything from months to decades and hence can mimic the timing variations caused by a planet. Furthermore, eccentricities this low are hard to detect by other means (e.g. spectroscopy), making it hard to rule them out, even for post common envelope systems.

Apsidal motion can be detected, or ruled out, using observations of the secondary eclipse (the transit of the white dwarf across the face of the M dwarf). This is because the precession of the apses causes the secondary eclipse to move in the opposite sense to the primary eclipse. Hence if the secondary eclipse times follow the same trend as the primary eclipse times then we can rule out apsidal motion as the cause of the variations. Unfortunately, since the best primary eclipse timing data usually comes from systems with hot, dominant white dwarfs (hence very shallow secondary eclipses), secondary eclipse timings are difficult and only possible in a handful of systems.

One system where secondary eclipse timings are currently possible is NN\,Serpentis. NN\,Ser is an eclipsing binary with a period of 3.1 hours \citep{haefner89}, consisting of a hot, 57,000\,K white dwarf \citep{wood91} and a low mass (0.111\,\MSUN) main-sequence companion \citep{parsons10nn}. The extreme temperature of the white dwarf and the close proximity of the M dwarf causes a large reflection effect on one side of the M dwarf. This large reflection effect actually increases the depth of the secondary eclipse since the white dwarf transits across the bright, heated face of the M dwarf, making it suitable for timing studies. Furthermore, the deep primary eclipse has been used to measure times with precisions as low as 0.01 seconds \citep{parsons10}. These high-precision times have revealed complex variations in the eclipse arrival times too large to be caused by Applegate's mechanism \citep{brinkworth06}. \citet{beuermann10} found that the timing variations are consistent with the gravitational effects of a 2.2\,{\MJUP} planet in a 7.7 year orbit and 6.9\,{\MJUP} planet in a 15.5 year orbit around the binary. Recent dynamical studies have shown that such a system is stable over the current cooling age of the white dwarf \citep{beuermann13,marsh13}. NN\,Ser is currently the only system where subsequent timing measurements have remained consistent with predictions from planetary models \citep{marsh13}, albeit with a greatly reduced number of stable solutions.

NN\,Ser represents one of the most compelling cases for the existence of planets around an evolved binary. The planetary interpretation of the timing variations has proven predictive power and there are very few alternative mechanisms able to drive the timing variations. Here we present 15 secondary eclipse times spanning more than a decade and use them to test if the timing variations are caused by apsidal motion, one of the few possible alternative mechanisms.

\section{Observations and their Reduction}

The majority of the data presented in this paper were obtained with the high-speed camera ULTRACAM \citep{dhillon07}, mounted as a visitor instrument on the William Herschel Telescope (WHT), ESO Very Large Telescope (VLT) and ESO New Technology Telescope (NTT). ULTRACAM is a frame transfer camera which splits the incoming light into three beams; red, green and blue. The secondary eclipse in NN\,Ser is deeper at longer wavelengths \citep{parsons10nn}, hence we only use data taken from the red beam, equipped with either a $r'$ or $i'$ band filter, as only this data can provide reliable timings. 

We supplement our ULTRACAM data with high-speed $J$ band observations of NN\,Ser using the infrared imager HAWK-I on the VLT \citep{kissler08}. The reflection effect in NN\,Ser is strongest at near-infrared wavelengths, resulting in a deeper secondary eclipse, which compensates for the overall reduction in flux at these longer wavelengths (relative to the ULTRACAM optical observations). We used HAWK-I in fast photometry mode, allowing us to window the detector and reduce the deadtime between frames to a few microseconds. All data were de-biased (dark subtracted in the case of HAWK-I), flat-fielded and extracted using aperture photometry within the ULTRACAM pipeline \citep{dhillon07}.

We initially fitted all of the light curves individually using the light curve model developed for our previous analysis of NN\,Ser \citep{parsons10nn} and Levenberg--Marquardt minimisation, allowing only the mid-eclipse times to vary. We limited our out-of-eclipse data to one eclipse width before and after the eclipse itself in order to reduce the impact of the reflection effect on the timings. We used these first estimates of the eclipse times to combine light curves in order to create high signal-to-noise light curves in each band ($r'$, $i'$ and $J$). NN\,Ser is ideally suited to this approach since its light curve is very stable, having never shown a flare or variations in the shape and size of the reflection effect since our first ULTRACAM observations in 2002. 

We then fitted these high signal-to-noise light curves allowing for a linear slope in the data. We included a linear slope to try to account for the possibility of  equatorial heat transport effects on the M dwarf, which could move the peak of the reflection away from phase 0.5 introducing a slope around this phase. Although we do not detect this effect, the secondary eclipse is shallow enough that it can be affected by such effects. Therefore, we accepted the additional statistical scatter introduced by including a slope, in order to reduce the above potential systematics.

\begin{figure}
\begin{center}
 \includegraphics[width=0.99\columnwidth]{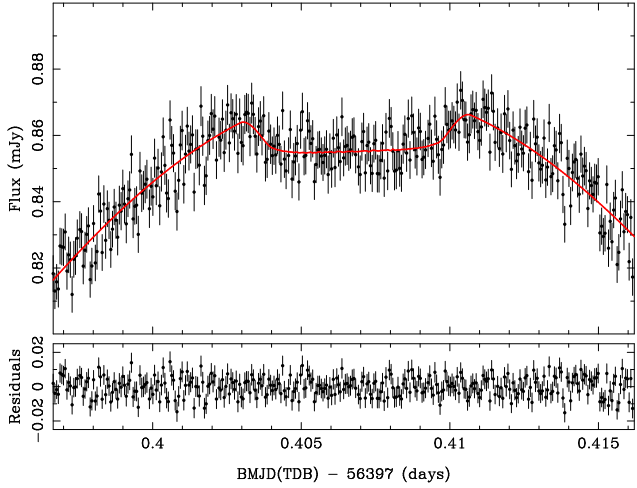}
 \caption{HAWK-I $J$ band secondary eclipse data (cycle 69598.5) with model fit (red line) overplotted. The lower panel shows the fit residuals}
 \label{fig:ecl_fit}
 \end{center}
\end{figure}

The resulting models (one for each band) were then used to refit the individual light curves allowing only the central eclipse time to vary and using Markov chain Monte Carlo (MCMC) minimisation. Fig.~\ref{fig:ecl_fit} shows an example of the model fit, and the fitted eclipse times are listed in Table~\ref{tab:etimes}. Our HAWK-I light curves showed some small scale correlated noise. We estimate what effect this had on the fits using the ``time-averaging'' method described in \citet{winn08}, whereby the residuals of the fit are binned over various timescales and the RMS recalculated. We found that the resulting $\beta$ parameter was small $<1.1$, and therefore any red noise has a very minor impact on the overall fits, not enough to affect our overall conclusions.

In poor conditions, such as those experienced in 2004 (cycle numbers around 44475), the secondary eclipse is often difficult to detect and leads to large uncertainties in the mid-eclipse times ($\sim$8 seconds). However, generally we are able to measure eclipse times to better than 4 seconds, and with our VLT+ULTRACAM data (cycle 53176.5) we reach a precision of 1.8 seconds.

\section{Results}

Fig.~\ref{fig:omc} shows the observed minus calculated secondary eclipse times for NN\,Ser relative to the planetary fits of \citet{marsh13}. We also show a vertically reflected version of the best planetary model. If the variations seen in the primary eclipse times were due to apsidal motion then the secondary eclipse times should follow this reflected trend. However, they show the same variations as the primary eclipse times, counter to what we would expect from apsidal motion. The most recent HAWK-I eclipse times are key to this, since they are more than 30 seconds (or $\sim$6$\sigma$) from this reflected trend.

The lower panel of Fig.~\ref{fig:omc} shows the residuals between the secondary eclipse times and the best planetary model. The secondary eclipse times do not completely agree with those of the primary eclipse, they generally appear slightly later than expected (i.e. they do not occur precisely at phase 0.5). A weighted mean of this offsets gives a delay of $3.3\pm1.0$\,seconds in the arrival time of the secondary eclipse.

\begin{table}
 \centering
  \caption{Secondary eclipse times for NN\,Ser}
  \label{tab:etimes}
  \begin{tabular}{@{}lllc@{}}
  \hline
Cycle   & BMJD(TDB)          & O-C   & Source            \\
No.     & (mid-eclipse)     & (sec) &                   \\
 \hline
38960.5 & 52412.012165(44) & 5.9   & WHT/ULTRACAM $r'$ \\
38967.5 & 52412.922698(53) & 3.5   & WHT/ULTRACAM $r'$ \\
38976.5 & 52414.093377(52) & -0.1  & WHT/ULTRACAM $r'$ \\
41797.5 & 52781.049453(46) & 3.5   & WHT/ULTRACAM $i'$ \\
41798.5 & 52781.179464(61) & -2.5  & WHT/ULTRACAM $i'$ \\
41805.5 & 52782.090016(50) & -3.1  & WHT/ULTRACAM $i'$ \\
41828.5 & 52785.081966(53) & 6.1   & WHT/ULTRACAM $i'$ \\
44472.5 & 53129.013865(95) & 12.6  & WHT/ULTRACAM $i'$ \\
44473.5 & 53129.143641(92) & -13.7 & WHT/ULTRACAM $i'$ \\
44480.5 & 53130.054368(95) & 0.6   & WHT/ULTRACAM $i'$ \\
49662.5 & 53804.129634(82) & 12.0  & WHT/ULTRACAM $r'$ \\
49663.5 & 53804.259706(53) & 11.3  & WHT/ULTRACAM $r'$ \\
53176.5 & 54261.231056(24) & 2.7   & VLT/ULTRACAM $i'$ \\
61226.5 & 55308.375950(68) & 4.1   & NTT/ULTRACAM $i'$ \\
69598.5 & 56397.407037(52) & 3.9   & VLT/HAWK-I $J$    \\
70456.5 & 56509.015866(56) & 5.3   & VLT/HAWK-I $J$    \\

\hline
\end{tabular}
\end{table}

\section{Discussion}

\subsection{No evidence of apsidal motion}

Eccentricity causes a shift in the secondary eclipse arrival times, relative the those of the primary eclipse, of the form \citep{barlow12}
\begin{equation}
\frac{2}{\pi} P e \cos{\omega}
\end{equation}
where $P$ is the period of the binary, $e$ is the eccentricity and $\omega$ is the angle of the apsides. For NN\,Ser this means that an eccentricity as low as $10^{-3}$ would cause variations in the eclipse times of around 10 seconds, a similar magnitude to the variations detected. 

The rate of secular apsidal motion ($\dot{\omega}$) is the sum of the tidal, rotational and relativistic terms. Ignoring the white dwarf's contribution (which is negligible compared to the M dwarf) these are
\begin{eqnarray}
\dot{\omega}_\mathrm{tide} & = & 15\Omega \left(\frac{R_\mathrm{dM}}{a}\right)^5 \frac{M_\mathrm{WD}}{M_\mathrm{dM}} \frac{1+1.5e^2 + 0.125e^4}{(1-e^2)^5} k_\mathrm{dM}\\
\dot{\omega}_\mathrm{rot} & = & \Omega \left(\frac{R_\mathrm{dM}}{a}\right)^5 \frac{M_\mathrm{WD}+M_\mathrm{dM}}{M_\mathrm{dM}} \frac{(\Omega_\mathrm{dM} / \Omega)^2}{(1-e^2)^2} k_\mathrm{dM}\\
\dot{\omega}_\mathrm{GR} & = & \Omega \left( \frac{3G}{c^2} \right) \frac{M_\mathrm{WD}+M_\mathrm{dM}}{a(1-e^2)},
\end{eqnarray}
where $\Omega=2\pi/P$, $M_\mathrm{WD}$ is the mass of the white dwarf, $M_\mathrm{dM}$ and $R_\mathrm{dM}$ are the mass and radius of the M dwarf, $a$ is the orbital separation, $k_\mathrm{dM}$ is the apsidal precession constant for the M dwarf \citep{claret91} and $\Omega_\mathrm{dM}$ is the rotational angular velocity of the M dwarf.

\begin{figure*}
\begin{center}
 \includegraphics[width=0.7\textwidth]{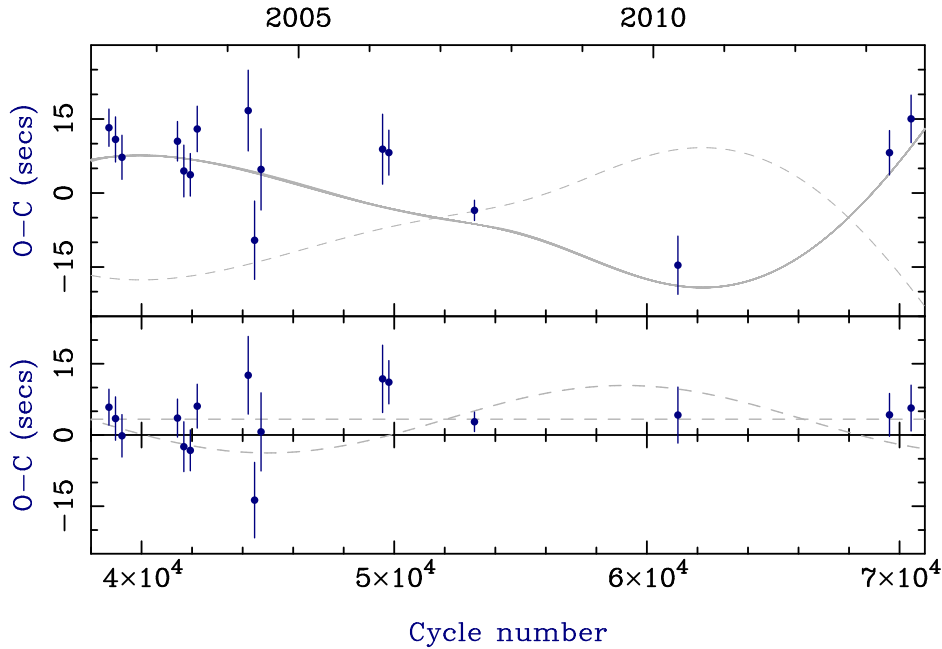}
 \caption{{\it Top panel:} observed minus calculated (O-C) eclipse arrival times for the secondary eclipses of NN\,Ser relative to the ephemeris BMJD(TDB)$=52942.934\,05+0.130\,080\,127 E$. Also shown are 50 best fit planetary models from \citet{marsh13} where the eccentricities of both planets were allowed to vary. The dashed line is a reflected version of the best planetary model. The secondary eclipse times would follow this reflected trend if the variations were due to apsidal motion. We have spread out observations taken close together to make them clearer (given the timing uncertainties this has no effect on the overall trend). {\it Bottom panel:} residuals of the secondary eclipse times relative to the best planetary model. The horizontal dashed line shows the mean value. The dashed sinusoid shows the variations we would expect if NN\,Ser had an eccentricity of $10^{-3}$, demonstrating that the actual eccentricity is unlikely to be larger than this value.}
 \label{fig:omc}
 \end{center}
\end{figure*}

Using the parameters from \citet{parsons10nn} and zero eccentricity (which gives an upper limit on the apsidal period) gives $\dot{\omega}_\mathrm{tide}\sim22^\circ/\mathrm{yr}$, $\dot{\omega}_\mathrm{rot}\sim9^\circ/\mathrm{yr}$ and $\dot{\omega}_\mathrm{GR}\sim5^\circ/\mathrm{yr}$. Hence $\dot{\omega}\sim36^\circ/\mathrm{yr}$ or 1 full cycle in $\sim10$ years, which is comparable to the timescale of the eclipse timing variations (even with an eccentricity as high as 0.1 the timescale remains similar). Any significant eccentricity would have been detected in our data, since it spans 11 years (i.e. at least one full apsidal cycle).

Given that such a small eccentricity (currently undetectable spectroscopically in NN\,Ser) can create timing variations of a similar amplitude and period to those seen in NN\,Ser means that it is important to rule out this mechanism if the planetary hypothesis is to survive. Therefore, our secondary eclipse timings, which follow the trend seen in the primary eclipse times and hence rule out apsidal motion, are an important result and, although not proving the planetary hypothesis, they certainly support it.

We place an upper limit on the eccentricity of NN\,Ser using the residuals of the secondary eclipse times relative to the planetary model (bottom panel of Fig.~\ref{fig:omc}). The sinusoid shows the signal we would expect if the binary had an eccentricity of $5\times10^{-4}$ (with an apsidal cycle time of 10 years), it is evident that a larger eccentricity (hence larger amplitude) is ruled out. Therefore, we limit the eccentricity of NN\,Ser to $e<10^{-3}$.

\subsection{The delay in the secondary eclipse times}

As previously mentioned, the secondary eclipse times do not occur exactly at phase 0.5, but rather occur slightly later than expected, by $3.3\pm1.0$\,seconds. The finite speed of light causes a delay between the primary and secondary eclipses in binary systems with unequal mass ratios. This R{\o}mer delay is given by \citep{kaplan10}
\begin{equation}
\Delta T = \frac{P}{\pi c}(K_\mathrm{dM}-K_\mathrm{WD}),
\end{equation}
where $K_\mathrm{dM}$ and $K_\mathrm{WD}$ are the radial velocity semi-amplitudes of the M dwarf and white dwarf respectively. For the measured parameters of NN\,Ser this delay is $2.84\pm0.04$ seconds \citep{parsons10nn}, and hence our measured offset is consistent with this effect. R{\o}mer delays have been detected in the sdB+dM binary 2MASS\,J1938+4603 \citep{barlow12} and the white dwarf plus A star binary KOI-74 \citep{bloemen12}, but this is the first detection of this effect in a white dwarf plus M dwarf binary.

\section{Conclusions}

We have measured the mid-eclipse times for the secondary eclipse of the white dwarf plus main-sequence binary NN\,Ser spanning a time period of more than a decade. Our results show that the secondary eclipse arrival times display a similar trend to the arrival times of the primary eclipse, ruling out apsidal motion as a possible cause of the observed timing variations. Given that fluctuations in the gravitational quadrupolar moment of the M dwarf have also been ruled out \citep{brinkworth06}, we are left with no known alternative explanations as to the origin of these variations beyond the circumbinary planet hypothesis. This result, along with recent studies demonstrating that the proposed planetary system is stable and has some predictive power \citep{beuermann13,marsh13}, makes NN\,Ser the most convincing case for the existence of planets around an evolved binary. However, despite the lack of alternative explanations, the existence of these planets (and those thought to exist around other compact binaries) are yet to be proven. Independent evidence is still required, such as direct detection of the planets or evidence of $N$-body effects in the eclipse timings. 

The lack of any obvious sinusoidal variations between the primary and secondary eclipse times of NN\,Ser limit its eccentricity to $e<10^{-3}$. We have also detected a delay in the arrival times of the secondary eclipses, relative to the primary eclipse times, of $3.3\pm1.0$ seconds. This delay is consistent with the predicted R{\o}mer delay of the binary, which is $2.84\pm0.04$ seconds.

\section*{Acknowledgments}

SGP acknowledges support from the Joint Committee ESO-Government of Chile. TRM and EB were supported under a grant from the UK's Science and Technology Facilities Council (STFC), ST/F002599/1. SPL and VSD were also supported by the STFC. MRS is supported by the Milenium Science Initiative, Chilean Ministry of Economy, Nucleus P10-022-F. CC acknowledges the support from ALMA-CONICYT Fund through grant 31100025. The results presented in this paper are based on observations collected at the European Southern Observatory under programme IDs 090.D-0384 and 091.D-0444

\bibliographystyle{mn_new}
\bibliography{eclipsers}

\begin{thebibliography}{34}
\expandafter\ifx\csname natexlab\endcsname\relax\def\natexlab#1{#1}\fi

\bibitem[{{Applegate}(1992)}]{applegate92}
{Applegate}, J.~H., 1992, \apj, 385, 621

\bibitem[{{Barlow} et~al.(2012){Barlow}, {Wade}, \& {Liss}}]{barlow12}
{Barlow}, B.~N., {Wade}, R.~A., {Liss}, S.~E., 2012, \apj, 753, 101

\bibitem[{{Beuermann} et~al.(2013){Beuermann}, {Dreizler}, \&
  {Hessman}}]{beuermann13}
{Beuermann}, K., {Dreizler}, S., {Hessman}, F.~V., 2013, \aap, 555, A133

\bibitem[{{Beuermann} et~al.(2010)}]{beuermann10}
{Beuermann}, K., et~al., 2010, \aap, 521, L60

\bibitem[{{Beuermann} et~al.(2011)}]{beuermann11}
{Beuermann}, K., et~al., 2011, \aap, 526, A53

\bibitem[{{Beuermann} et~al.(2012)}]{beuermann12}
{Beuermann}, K., et~al., 2012, \aap, 540, A8

\bibitem[{{Bloemen} et~al.(2012)}]{bloemen12}
{Bloemen}, S., et~al., 2012, \mnras, 422, 2600

\bibitem[{{Brinkworth} et~al.(2006){Brinkworth}, {Marsh}, {Dhillon}, \&
  {Knigge}}]{brinkworth06}
{Brinkworth}, C.~S., {Marsh}, T.~R., {Dhillon}, V.~S., {Knigge}, C., 2006,
  \mnras, 365, 287

\bibitem[{{Chauvin} et~al.(2005){Chauvin}, {Lagrange}, {Dumas}, {Zuckerman},
  {Mouillet}, {Song}, {Beuzit}, \& {Lowrance}}]{chauvin05}
{Chauvin}, G., {Lagrange}, A.-M., {Dumas}, C., {Zuckerman}, B., {Mouillet}, D.,
  {Song}, I., {Beuzit}, J.-L., {Lowrance}, P., 2005, \aap, 438, L25

\bibitem[{{Claret} \& {Gimenez}(1991)}]{claret91}
{Claret}, A., {Gimenez}, A., 1991, \aaps, 87, 507

\bibitem[{{Collier Cameron} et~al.(2010)}]{collier10}
{Collier Cameron}, A., et~al., 2010, \mnras, 407, 507

\bibitem[{{Dhillon} et~al.(2007)}]{dhillon07}
{Dhillon}, V.~S., et~al., 2007, \mnras, 378, 825

\bibitem[{{D{\"o}llinger} et~al.(2011)}]{dollinger11}
{D{\"o}llinger}, M.~P., et~al., 2011, in {Schuh}, S., {Drechsel}, H., {Heber},
  U., eds., American Institute of Physics Conference Series, vol. 1331 of
  \emph{American Institute of Physics Conference Series}, p.~79

\bibitem[{{Doyle} et~al.(2011)}]{doyle11}
{Doyle}, L.~R., et~al., 2011, Science, 333, 1602

\bibitem[{{G{\"a}nsicke} et~al.(2012){G{\"a}nsicke}, {Koester}, {Farihi},
  {Girven}, {Parsons}, \& {Breedt}}]{gansicke12}
{G{\"a}nsicke}, B.~T., {Koester}, D., {Farihi}, J., {Girven}, J., {Parsons},
  S.~G., {Breedt}, E., 2012, \mnras, 424, 333

\bibitem[{{Haefner}(1989)}]{haefner89}
{Haefner}, R., 1989, \aap, 213, L15

\bibitem[{{Han} et~al.(2013)}]{han13}
{Han}, C., et~al., 2013, \apj, 778, 38

\bibitem[{{Herrero} et~al.(2011){Herrero}, {Morales}, {Ribas}, \&
  {Naves}}]{herrero11}
{Herrero}, E., {Morales}, J.~C., {Ribas}, I., {Naves}, R., 2011, \aap, 526, L10

\bibitem[{{Hinse} et~al.(2012){Hinse}, {Go{\'z}dziewski}, {Lee},
  {Haghighipour}, \& {Lee}}]{hinse12}
{Hinse}, T.~C., {Go{\'z}dziewski}, K., {Lee}, J.~W., {Haghighipour}, N., {Lee},
  C.-U., 2012, \aj, 144, 34

\bibitem[{{Horner} et~al.(2013){Horner}, {Wittenmyer}, {Hinse}, {Marshall},
  {Mustill}, \& {Tinney}}]{horner13}
{Horner}, J., {Wittenmyer}, R.~A., {Hinse}, T.~C., {Marshall}, J.~P.,
  {Mustill}, A.~J., {Tinney}, C.~G., 2013, \mnras, 435, 2033

\bibitem[{{Kaplan}(2010)}]{kaplan10}
{Kaplan}, D.~L., 2010, \apjl, 717, L108

\bibitem[{{Kissler-Patig} et~al.(2008)}]{kissler08}
{Kissler-Patig}, M., et~al., 2008, \aap, 491, 941

\bibitem[{{Lee} et~al.(2009){Lee}, {Kim}, {Kim}, {Koch}, {Lee}, {Kim}, \&
  {Park}}]{lee09}
{Lee}, J.~W., {Kim}, S.-L., {Kim}, C.-H., {Koch}, R.~H., {Lee}, C.-U., {Kim},
  H.-I., {Park}, J.-H., 2009, \aj, 137, 3181

\bibitem[{{Marsh} et~al.(2013)}]{marsh13}
{Marsh}, T.~R., et~al., 2013, \mnras

\bibitem[{{Mustill} et~al.(2013){Mustill}, {Marshall}, {Villaver}, {Veras},
  {Davis}, {Horner}, \& {Wittenmyer}}]{mustill13}
{Mustill}, A.~J., {Marshall}, J.~P., {Villaver}, E., {Veras}, D., {Davis},
  P.~J., {Horner}, J., {Wittenmyer}, R.~A., 2013, \mnras

\bibitem[{{Parsons} et~al.(2010{\natexlab{a}}){Parsons}, {Marsh},
  {Copperwheat}, {Dhillon}, {Littlefair}, {G{\"a}nsicke}, \&
  {Hickman}}]{parsons10nn}
{Parsons}, S.~G., {Marsh}, T.~R., {Copperwheat}, C.~M., {Dhillon}, V.~S.,
  {Littlefair}, S.~P., {G{\"a}nsicke}, B.~T., {Hickman}, R.,
  2010{\natexlab{a}}, \mnras, 402, 2591

\bibitem[{{Parsons} et~al.(2010{\natexlab{b}})}]{parsons10}
{Parsons}, S.~G., et~al., 2010{\natexlab{b}}, \mnras, 407, 2362

\bibitem[{{Potter} et~al.(2011)}]{potter11}
{Potter}, S.~B., et~al., 2011, \mnras, 416, 2202

\bibitem[{{Veras} \& {Tout}(2012)}]{veras12}
{Veras}, D., {Tout}, C.~A., 2012, \mnras, 422, 1648

\bibitem[{{Welsh} et~al.(2012)}]{welsh12}
{Welsh}, W.~F., et~al., 2012, \nat, 481, 475

\bibitem[{{Winn} et~al.(2008)}]{winn08}
{Winn}, J.~N., et~al., 2008, \apj, 683, 1076

\bibitem[{{Wolszczan} \& {Frail}(1992)}]{wolszcan92}
{Wolszczan}, A., {Frail}, D.~A., 1992, \nat, 355, 145

\bibitem[{{Wood} \& {Marsh}(1991)}]{wood91}
{Wood}, J.~H., {Marsh}, T.~R., 1991, \apj, 381, 551

\bibitem[{{Zorotovic} \& {Schreiber}(2013)}]{zorotovic13}
{Zorotovic}, M., {Schreiber}, M.~R., 2013, \aap, 549, A95

\end{thebibliography}

\label{lastpage}

\end{document}